\documentstyle[preprint,floats,pra,eqsecnum,aps,psfig]{revtex}
\tightenlines
\begin{document} \draft

\title{High-speed Contraction of Transverse Rotations to\\
Gauge Transformations\thanks{presented at the International
Workshop on Quantum Groups, Deformations, and Contractions,
Istanbul, Turkey (September 1997); to be published in the
proceedings.}}

\author{Y. S. Kim\thanks{e-mail: yskim@physics.umd.edu}}
\address{Department of Physics, University of Maryland, College Park,
Maryland 20742, U.S.A.}

\maketitle

\begin{abstract}
The In\"on\"u-Wigner contraction is applied to special relativity
and the little groups of the Lorentz group.  If the $O(3)$ symmetry
group for massive particle is boosted to an infinite-momentum frame,
it becomes contracted to a combination of the cylindrical group and
the two-dimensional Euclidean group.  The Euclidean component becomes
the Lorentz condition applicable to the electromagnetic four-potential,
and the cylindrical component leads to the helicity and gauge degrees
of freedom.  The rotation around the cylindrical axis corresponds to
the helicity, while the translation parallel to the axis on the
cylindrical surface leads to a gauge transformation.

\end{abstract}

\section{Introduction}\label{intro}
In his 1939 paper on representations of the Poincar\'e group,
Wigner formulated the internal space-time symmetries of relativistic
particles in terms of the little groups whose transformations leave
the four-momentum of a given free particle invariant~\cite{wig39}.
He showed that the little groups for massive and massless particles
are isomorphic respectively to $O(3)$ (three-dimensional rotation group)
and $E(2)$ (two-dimensional Euclidean group).

In 1953, In\"on\"u and Wigner introduced to physics the concept and
techniques of group contractions~\cite{inonu53}, and they studied
the contraction of $O(3)$ to $E(2)$ in detail.   They also pointed out
that the Lorentz group can be contracted to the Galilean group in the
low-speed limit.
In 1968~\cite{bacryc68}, Bacry and Chang suggested the idea that the
high-speed limit of the Poincar\'e group could simplify quantum field
theory in the infinite-momentum frame~\cite{wein66}.  They then boosted
all ten generators of the Poincar\'e group and observed that the
contracted group has only seven generators.  Since Wigner's little
groups are subgroups of the Poincar\'e group, the Bacry-Chang procedure
contains the contractions of those little groups.

In this report, we work out the high-speed contraction of the
$O(3)$-like little group in terms of ellipsoidal deformations of a
sphere in the three-dimensional space, namely (American) football-like
and pancake-like deformations. We shall
see that both deformations are needed for the description of the
high-speed limiting process.  It will be shown that the football-like
and pancake-like deformations lead to a cylindrical surface and a flat
plane respectively.  The flat-plane deformation leads to the contraction
to the two-dimensional Euclidean group, while the football-like
deformation leads to a cylindrical surface.

It is shown further that the flat-surface collapse corresponds to the
Lorentz condition applicable to four-potentials for massless particles.
Rotations around the cylindrical axis correspond to the helicity.  It
is shown also that the translation along the axis on the cylindrical
surface corresponds to a gauge transformation.

In Sec.~\ref{contrac}, we explain what group contraction is in terms of
the north-pole approximation of a spherical surface, and how it can
it can be applied to contractions of the Lorentz group in the low-speed
limit.
In Sec.~\ref{review}, we review what Wigner did in his 1939 paper and
what have been done since then.  This paper has a stormy history
because it did not explain where Maxwell's covariant theory stands
within his representation scheme.  We now have a clear resolution of
this question.
In Sec.~\ref{3dim}, we study the three-dimensional rotation group
and its contractions to the cylindrical and two-dimensional Euclidean
groups.  It is shown that both of these contractions can be combined
into a single four-by-four representation.  In this section, the
four-dimensional group theoretical language of the little groups is
translated into a three-dimensional geometrical language based on
spheres, ellipsoids, cylinders, and flat surfaces.

In Sec.~\ref{cone}, the generators of the little groups are given
the light-cone coordinate system.  It is shown that these generators
are identical with the combined geometry of the cylindrical group and
the Euclidean group discussed in Sec.~\ref{3dim}.  The geometry of
Sec.~\ref{3dim} therefore gives a comprehensive description of the
little groups for massive and massless particles.
In Sec.~\ref{cyl}, we study how the four-vectors become contracted
in the high-speed limit and study also how the cylindrical group
describes the helicity and gauge degrees of freedom for massless
particles.
In Sec.~\ref{lboost}, the Lorentz boost is applied directly to the
transformation matrices.  In this way, we present a direct proof
that transverse rotations become contracted to gauge transformations.

\section{What is the Group Contraction?}\label{contrac}
If there is a curved line on a plane, we can define a straight line
tangent to the curve at a given point.  This led to the branch
of mathematics known today as differential calculus.  Likewise, if
there is a curved surface in a three-dimensional space, we can define
a flat plane tangent to the surface.  We live and work on a sphere
with a large radius called the earth, but the diameter of our city
is much smaller than the radius of the earth.  For this reason, the
city appears to be on a flat plane, which is tangent to the spherical
surface of the earth.

We are accustomed to associate the spherical surface with the
three-dimensional rotation group or $O(3)$, but are not accustomed to
think the flat plane in terms of the two-dimensional Euclidean group.
On the other hand, it is not difficult to construct a group theory
of transformations on this plane.  It is commonly known as the
two-dimensional Euclidean group or $E(2)$.  With this point in mind,
In\"on\"u and Wigner in 1953 introduced to physics the concept
and technique of group contractions~\cite{inonu53}.   They started
with the set of generators and the closed Lie algebra for a given
group.  They then considered the case when some of the generators
change the form with a scale change.  For instance, the generators
of the three-dimensional rotation group are
\begin{equation}
L_{1} = -i\left(y {\partial \over \partial z} -
                 z {\partial \over \partial y}\right) , \quad
L_{2} = -i\left(z {\partial \over \partial x} -
                 x {\partial \over \partial z}\right) , \quad
L_{3} = -i\left(x {\partial \over \partial y} -
                 y {\partial \over \partial x}\right) .
\end{equation}
They of course satisfy the commutation relations
\begin{equation}\label{o3com}
\left[L_{i}, L_{j}\right] = i \epsilon_{ijk} J_{k} .
\end{equation}
Let us next consider the sphere with a large radius.  If we consider
a small region on the north pole, the $z$ variable can be replaced
by $R$, which is the radius of the sphere and is much larger than
$x$ and $y$ variables.  Then, $L_{3}$ remains unchanged, but
$L_{1}$ and $L_{2}$ become
\begin{equation}
L_{1} = iR {\partial \over \partial y} = -RP_{2} , \qquad
L_{2} = -iR {\partial \over \partial x} = RP_{1} ,
\end{equation}
where $P_{1}$ and $P_{2}$ are generators of translations along the
$x$ and $y$ directions respectively.  They can also be written as
\begin{equation}
P_{1} = -i {\partial \over \partial x} = {1 \over R} L_{2} , \qquad
P_{2} = -i {\partial \over \partial y} = -{1 \over R} L_{1} .
\end{equation}
This leads to the commutation relations
\begin{equation}
\left[P_{1}, L_{3}\right] = -iP_{2} ,  \qquad
\left[P_{2}, L_{3}\right] = iP_{1} ,
\end{equation}
and
\begin{equation}
\left[P_{1}, P_{2}\right] = i \left({1 \over R}\right)^{2} J_{3} .
\end{equation}
The right hand side of this expression becomes zero when R becomes
large.  We then end up with a closed set of three commutation
relations:
\begin{equation}\label{e2com}
\left[P_{1}, P_{2}\right] = 0 ,  \qquad
\left[P_{1}, L_{3}\right] = -iP_{2} ,  \qquad
\left[P_{2}, L_{3}\right] = iP_{1} .
\end{equation}
These three operators generate two translations and rotations around
the origin on the $xy$ plane.  They generate the $E(2)$ group.

In order to see the contraction of the polar variables
$(R, \theta, \phi)$, let us write $(x, y, z)$ as
\begin{equation}
x = R sin\theta\cos\phi , \quad  y = R \sin\theta\sin\phi ,
\quad z = R \cos\theta .
\end{equation}
If $R$ becomes infinite while $x$ and $y$ remain finite, the
angle $\theta$ has to be small, and we can write $(z, y, z)$ as
\begin{equation}
x = R \theta\cos\phi , \quad  y = R \theta\sin\phi ,
\quad z = R ,
\end{equation}
with a finite value of $R\theta$.  This is how the angle variables
$\theta$ and $\phi$ are transformed into $x$ and $y$.

We have illustrated above how $O(3)$ is contracted to $E(2)$ through
a flat surface tangent to a sphere.  A curved surface does not have
to be a spherical.  Instead of the spherical surface, we can consider
a plane tangent to a hyperboloidal surface at its top or
bottom~\cite{gilmore74}.  This is the contraction of $O(2,1)$ to
$E(2)$.  The tangential surface does not have to be a plane.  For the
case of sphere, we can construct a cylinder which is tangential to the
equatorial belt of the sphere.  If we can construct a group theory on
the surface of the cylinder, it is possible to contract $O(3)$ to the
cylindrical group~\cite{kiwi87jm}.

In their 1953 paper, In\"on\"u and Wigner were primarily interested
in contractions of representations within the framework of unitary
representations.  On the other hand, they also gave an example based
on finite-dimensional matrix representations.  They considered the
contraction of the Lorentz group into the Galilean group for infinite
value of the speed of light.  Let us review their reasoning.  If the
Lorentz boost is made in the $x$ direction, the transformation
matrix is
\begin{equation}
\pmatrix{x' \cr ct'}  = \pmatrix{\cosh\lambda & \sinh\lambda \cr
\sinh\lambda & \cosh\lambda} \pmatrix{x \cr ct},
\end{equation}
where $c$ is the velocity of light.  Now the space-time column vector
can be written as
\begin{equation}
\pmatrix{x \cr ct}  = \pmatrix{1 & 0 \cr 0 & c} \pmatrix{x \cr t} .
\end{equation}
Thus,
\begin{equation}
\pmatrix{x' \cr t'}  = \pmatrix{\cosh\lambda & c (\sinh\lambda) \cr
(\sinh\lambda)/c  & \cosh\lambda} \pmatrix{x \cr t} .
\end{equation}
Let us introduce a variable $v$, where
\begin{equation}
{v \over c} = \sinh\lambda .
\end{equation}
If $c$ becomes infinite while $v$ remains finite, $\lambda$ has to be
vanishingly small, and the transformation matrix becomes
\begin{equation}
\pmatrix{x' \cr t'}  = \pmatrix{1 & v \cr 0 & 1} \pmatrix{x \cr t} ,
\end{equation}
resulting in
\begin{equation}
x' = x + vt, \qquad t' = t .
\end{equation}
This is how the Lorentz group becomes contracted to the Galilean
group.

In Sec. \ref{3dim} of this paper, we shall use the three-by-three
matrices to contract $O(3)$ to $E(2)$ and to the cylindrical group.
We then combine these two contractions to construct the high-speed
contraction of the $O(3)$-like little group to the $E(2)$-like little
group.  Before doing this let us review the history of Wigner's
little groups.

\section{Historical Review of Wigner's Little Groups}\label{review}
In 1939, Wigner observed that internal space-time symmetries of
relativistic particles are dictated by their respective little
groups~\cite{wig39}.  The little group is the maximal subgroup of the
Lorentz group which leaves the four-momentum of the particle invariant.
The Lorentz group is generated by three rotation generators $J_{i}$
and three boost generators $K_{i}$.  If a massive particle is at
rest, it momentum is invariant under three-dimensional rotations.
Thus, its little group is generated by $J_{1}, J_{2}$, and $J_{3}$,
and its spin orientation is changed under the little group
transformation.

For a massless particle, it is not possible to find a Lorentz frame
in which the particle is at rest.  We can however assume that its
momentum is in the $z$ direction.  Then the momentum is invariant
under the subgroup of the Lorentz group generated by
\begin{equation}
J_{3} , \qquad N_{1} = K_{1} - J_{2} , \qquad N_{2} = K_{2} + J_{1} .
\end{equation}
Wigner noted in his 1939 paper that these generators satisfy the
same set of commutation relations as those for the two-dimensional
Euclidean group consisting of one rotation and translations in two
different directions.

The 1939 paper indeed gives a covariant picture of massive particles
with spins, and connects the helicity of massless particle
with the rotational degree of freedom in the group $E(2)$.  This
paper also gives many homework problems for us to solve.

\begin{itemize}
\item First, like the three-dimensional rotation group, $E(2)$ is a
three-parameter group.  It contains two translational degrees of freedom
in addition to the rotation.  What physics is associated with the
translational-like degrees of freedom for the case of the $E(2)$-like
little group?

\item Second, as is shown by In\"on\"u and Wigner~\cite{inonu53}, the
rotation group $O(3)$ can be contracted to $E(2)$.  Does this mean that
the $O(3)$-like little group can become the $E(2)$-like little group in a
certain limit?

\item Third, it is possible to interpret the Dirac equation in terms of
Wigner's representation theory~\cite{barg48}.  Then, why is it not
possible to find a place for Maxwell's equations in the same theory?

\item Fourth, the proton was found to have a finite space-time extension
in 1955~\cite{hofsta55}, and the quark model has been established in
1964~\cite{gell64}.  The concept of relativistic extended particles has
now been firmly established.  Is it then possible to construct a
representation of the Poincar\'e group for particles with space-time
extensions?
\end{itemize}

The list could be endless, but let us concentrate on the above four
questions.  As for the first question, it has been shown by various
authors that the translation-like degrees of freedom in the $E(2)$-like
little group is the gauge degree of freedom for massless
particles~\cite{janner71,kuper76}.  As for the second question, it is
not difficult to guess that the $O(3)$-like little group becomes the
$E(2)$-like little group in the limit of large
momentum/mass~\cite{bacryc68,misra76}.  However, the non-trivial
result is that the transverse rotational degrees of freedom become
contracted to the gauge degree of freedom~\cite{hks83}.

Then there comes the third question.  Indeed, in 1964~\cite{wein64},
Weinberg found a place for the electromagnetic tensor in Wigner's
representation theory.  He accomplished this by constructing from the
$SL(2,c)$ spinors all the representations of massless fields which are
invariant under the translation-like transformations of the $E(2)$-like
little group.  Since the translation-like transformations are gauge
transformations, and since the electromagnetic tensor is gauge-invariant,
Weinberg's construction should contain the electric and magnetic fields,
and it indeed does~\cite{baskal97}.

Next question is whether it is possible to construct electromagnetic
four-potentials.  After identifying the translation-like degrees of
freedom as gauge degrees of freedom, this becomes a tractable problem.
It is indeed possible to construct gauge-dependent four-potentials
from the $SL(2,c)$ spinors~\cite{baskal97,hks86}.  Yes, both the field
tensor and four-potential now have their proper places in Wigner's
representation theory.  The Maxwell theory and the Poincar\'e group
are perfectly consistent with each other.

The fourth question is about whether Wigner's little groups are
applicable to high-energy particle physics where accelerators produce
Lorentz-boosted extended hadrons such as high-energy protons.  The
question is whether it is possible to construct a representation of the
Poincar\'e group for hadrons which are believed to be bound states of
quarks~\cite{knp86,fkr71}.  This representation should describe
Lorentz-boosted hadrons.  Next question then is whether those boosted
hadrons give a description of Feynman's parton picture~\cite{fey69}
in the limit of large momentum/mass.  These issues have also been
discussed in the literature~\cite{knp86,kn77a}.

The author of this report was fortunate enough to have a close
collaboration with Professor Wigner in his late years.  Sections
\ref{3dim} and \ref{cone} of the present report are based on the
1990 joint paper by Wigner and the author~\cite{kiwi90jm},
where the little groups were translated into a geometrical language.

\section{Three-dimensional Geometry of the Little Groups}\label{3dim}
The little groups for massive and massless particles are isomorphic
to $O(3)$ and $E(2)$ respectively.  It is not difficult to construct
the $O(3)$-like geometry of the little group for a massive particle at
rest~\cite{wig39}.  In the three-by-three matrix representation, the
generators applicable to the column vector $(x, y, z)$ are
\begin{equation}\label{o3gen}
L_{1} = \pmatrix{0&0&0\cr0&0&-i\cr0&i&0} , \quad
L_{2} = \pmatrix{0&0&i\cr0&0&0\cr-i&0&0} , \quad
L_{3} = \pmatrix{0&-i&0\cr i &0&0\cr0&0&0} .
\end{equation}
These matrices satisfy the three commutation relations given in
Eq.(\ref{o3com}).  The Euclidean group $E(2)$ is generated by
$L_{3}, P_{1}$ and $P_{2}$, with
\begin{equation}
P_{1} = \pmatrix{0&0&i\cr0&0&0\cr0&0&0} , \qquad
P_{2} = \pmatrix{0&0&0\cr0&0&i\cr0&0&0} ,
\end{equation}
and they satisfy the commutation relations given in Eq.(\ref{e2com})
for the $E(2)$ group.  The generator $L_{3}$ is given in
Eq.(\ref{o3gen}).  When applied to the vector space $(x, y, 1)$,
$P_{1}$ and $P_{2}$ generate translations on in the $xy$ plane.
The geometry of $E(2)$ transformations is quite familiar to our
daily life.

Let us transpose the above algebra.  Then $P_{1}$ and $P_{2}$ become
$Q_{1}$ and $Q_{2}$ respectively, where
\begin{equation}
Q_{1} = \pmatrix{0&0&0\cr0&0&0\cr i &0&0} , \qquad
Q_{2} = \pmatrix{0&0&0\cr0&0&0\cr0&i&0} ,
\end{equation}
respectively.  Together with $L_{3}$, these generators satisfy the same
set of commutation relations as that for $L_{3}, P_{1}$, and $P_{2}$
given in Eq.(\ref{e2com}):
\begin{equation}
[Q_{1}, Q_{2}] = 0 , \qquad [L_{3}, Q_{1}] = iQ_{2} , \qquad
[L_{3}, Q_{2}] = -iQ_{1} .
\end{equation}
These matrices generate transformations of a point on a circular
cylinder.  Rotations around the cylindrical axis are generated
by $L_{3}$.  The $Q_{1}$ and $Q_{2}$ matrices generate the
transformation:
\begin{equation}\label{cyltrans}
\exp{\left(-i\xi Q_{1} - i\eta Q_{2}\right)} =
\pmatrix{1&0&0 \cr 0&1&0 \cr \xi & \eta & 1} .
\end{equation}
When applied to the space $(x, y, z)$, this matrix changes the value of
$z$ while leaving the $x$ and $y$ variables invariant~\cite{kiwi87jm}.
This corresponds to a translation along the cylindrical axis.  We shall
call the group generated by $L_{3}, Q_{1}$ and $Q_{2}$ the cylindrical
group~\cite{kiwi87jm}.

We can achieve the contractions to the Euclidean and to the cylindrical
groups by taking the large-radius limits of
\begin{equation}
P_{1} = {1\over R} B^{-1} L_{2} B ,
\qquad P_{2} = -{1\over R} B^{-1} L_{1} B ,
\end{equation}
and
\begin{equation}
Q_{1} = -{1\over R}B L_{2}B^{-1} , \qquad
Q_{2} = {1\over R} B L_{1} B^{-1} ,
\end{equation}
where
\begin{equation}\label{br3}
B(R) = \pmatrix{1&0&0\cr0&1&0\cr0&0&R}  .
\end{equation}
The vector spaces to which the above generators are applicable are
$(x, y, z/R)$ and $(x, y, Rz)$ for the Euclidean and cylindrical groups
respectively.  They can be regarded as the north-pole and equatorial-belt
approximations of the spherical surface respectively.

Next, let us consider linear combinations:
\begin{equation}
F_{1} = P_{1} + Q_{1} , \qquad F_{2} = P_{2} + Q_{2} .
\end{equation}
Since $P_{1} (P_{2})$ commutes with $Q_{2} (Q_{1})$, these operators
satisfy commutation relations:
\begin{equation}\label{commuf}
[F_{1}, F_{2}] = 0 , \qquad [L_{3}, F_{1}] = iF_{2} , \qquad
[L_{3}, F_{2}] = -iF_{1} .
\end{equation}
Indeed, this is another set of the $E(2)$-like commutation relations.
However, we cannot make this addition using the three-by-three matrices
for $P_{i}$ and $Q_{i}$ to construct three-by-three matrices for $F_{1}$
and $F_{2}$, because the vector spaces are different for the $P_{i}$ and
$Q_{i}$ representations.  We can accommodate this difference by creating
two different $z$ coordinates, one with a contracted $z$ and the other
with an expanded $z$, namely $(x, y, Rz, z/R)$.  Then the generators
become four-by-four matrices, and $F_{1}$ and $F_{2}$ take the form
\begin{equation}\label{f1f2}
F_{1} = \pmatrix{0&0&0&i \cr 0&0&0&0 \cr i&0&0&0 \cr 0&0&0&0} , \qquad
F_{2} = \pmatrix{0&0&0&0 \cr 0&0&0&i \cr 0&i&0&0 \cr 0&0&0&0} .
\end{equation}
The rotation generator $L_{3}$ is also a four-by-four matrix:
\begin{equation}\label{2rot}
L_{3} = \pmatrix{0&-i&0&0 \cr i&0&0&0 \cr 0&0&0&0 \cr 0&0&0&0} .
\end{equation}
These four-by-four matrices satisfy the $E(2)$-like commutation
relations of Eq.(\ref{commuf}).  The $B(R)$ matrix of Eq.(\ref{br3})
can now be combined with its inverse to become a four-by-four matrix:
\begin{equation}\label{br4}
B(R) = \pmatrix{1&0&0&0 \cr 0&1&0&0 \cr 0&0&R&0 \cr 0&0&0&1/R} .
\end{equation}

Next, let us consider the transformation matrix generated by the above
matrices.  It is easy to visualize the transformations generated by
$P_{i}$ and $Q_{i}$.  It would be easy to visualize the transformation
generated by $F_{1}$ and $F_{2}$, if $P_{i}$ commuted with $Q_{i}$.
However, $P_{i}$ and $Q_{i}$ do not commute with each other, and the
transformation matrix takes a somewhat complicated form:
\begin{equation}\label{compli}
\exp{\left(-i\xi F_{1} - i\eta F_{2}\right)} =
\pmatrix{1 & 0 & 0 & \xi \cr 0 & 1 & 0 & \eta \cr
\xi & \eta & 1 & (\xi ^{2} + \eta ^{2})/2 \cr 0 & 0 & 0 & 1}  .
\end{equation}
This matrix performs both the Euclidean and cylindrical transformations.
We shall see what role this matrix plays in special relativity and
the Lorentz group.

\section{Little Groups in the Light-cone Coordinate System}\label{cone}
Let us now study the group of Lorentz transformations using the
light-cone coordinate system.  If the space-time coordinate is
specified by $(x, y, z, t)$, then the light-cone coordinate variables
are $(x, y, u, v)$ for a particle moving in the $z$ direction, where
\begin{equation}
u = (z + t)/\sqrt{2} , \qquad v = (t - z)/\sqrt{2} .
\end{equation}
The transformation from the conventional space-time coordinate to the
above system is achieved through the coordinate transformation
\begin{equation}\label{simil}
S = \pmatrix{1&0&0&0 \cr 0&1&0&0 \cr
0&0&1/\sqrt{2} & 1/\sqrt{2} \cr 0 & 0&-1/\sqrt{2} & 1/\sqrt{2}} .
\end{equation}
In the light-cone coordinate system, the generators of Lorentz
transformations are
\begin{eqnarray}
J_{1} &=& {1 \over \sqrt{2}} \pmatrix{0&0&0&0 \cr 0&0&-i&i \cr0&i&0&0
\cr0&-i&0&0} , \qquad
K_{1} = {1 \over \sqrt{2}} \pmatrix{0&0&i&i \cr 0&0&0&0 \cr i&0&0&0
\cr i&0&0&0} ,\nonumber \\[2mm]
J_{2} &=& {1\over\sqrt{2}}\pmatrix{0&0&i&-i \cr 0&0&0&0 \cr -i&0&0&0
\cr i&0&0&0} , \qquad
K_{2} = {1\over\sqrt{2}}\pmatrix{0&0&0&0 \cr 0&0&i&i \cr 0&i&0&0
\cr0&i&0&0} , \nonumber \\[2mm]
J_{3} &=& \pmatrix{0&-i&0&0 \cr i&0&0&0 \cr 0&0&0&0 \cr 0&0&0&0} ,
\qquad K_{3} = \pmatrix{0&0&0&0 \cr 0&0&0&0 \cr 0&0&i&0 \cr 0&0&0&-i} .
\end{eqnarray}
where $J_{1}, J_{2}$, and $J_{3}$ are the rotation generators, and
$K_{1}, K_{2}$, and $K_{3}$ are the generators of boosts along the
three orthogonal directions.

If a massive particle is at rest, its little group is generated by
$J_{1}, J_{2}$ and $J_{3}$.  For a massless particle moving in the
$z$ direction, the little group is generated by $N_{1}, N_{2}$ and
$J_{3}$, where
\begin{equation}
N_{1} = K_{1} - J_{2} , \qquad N_{2} = K_{2} + J_{1} ,
\end{equation}
which can be written in the matrix form as
\begin{equation}
N_{1} = \sqrt{2} \pmatrix{0&0&0&i \cr 0&0&0&0 \cr i&0&0&0 \cr 0&0&0&0} ,
\qquad
N_{2} = \sqrt{2} \pmatrix{0&0&0&0 \cr 0&0&0&i \cr 0&i&0&0 \cr 0&0&0&0} .
\end{equation}
These matrices satisfy the commutation relations:
\begin{equation}\label{e2com2}
[J_{3}, N_{1}] =i N_{2} ,\qquad [J_{3}, N_{2}] = -i N_{1} , \qquad
[N_{1}, N_{2}] = 0 .
\end{equation}
Let us go back to $F_{1}$ and $F_{2}$ of Eq.(\ref{f1f2}).  Indeed, they
are proportional to $N_{1}$ and $N_{2}$ respectively:
\begin{equation}
N_{1} = \sqrt{2} F_{1} , \qquad N_{2} = \sqrt{2} F_{2} .
\end{equation}
Since $F_{1}$ and $F_{2}$ are somewhat simpler than $N_{1}$ and $N_{2}$,
and since the commutation relations of Eq.(\ref{e2com2}) are invariant
under multiplication of $N_{1}$ and $N_{2}$ by constant factors, we shall
hereafter use $F_{1}$ and $F_{2}$ for $N_{1}$ and $N_{2}$.

In the light-cone coordinate system, the boost matrix takes the form
\begin{equation}\label{boost}
B(R) = \exp \pmatrix{-i\rho K_{3}} =
\pmatrix{1&0&0&0 \cr 0&1&0&0 \cr 0&0&R&0 \cr 0&0&0&1/R} ,
\end{equation}
with $\rho = \ln(R)$, and
\begin{equation}
R = \left({1 + \beta \over 1 - \beta}\right)^{1/2} ,
\end{equation}
where $\beta$ is the velocity parameter of the particle.  The boost is
along the $z$ direction.  This boost matrix takes the same form as the
four-by-four contraction matrix given in Eq.(\ref{br4}).  Under this
transformation, $x$ and $y$ coordinates are invariant, and the
light-cone variables $u$ and $v$ are transformed as
\begin{equation}
u' = Ru , \qquad v' = v/R .
\end{equation}
If we boost $J_{2}$ and $J_{1}$ and divide them by $\sqrt{2}R$, as
$$
W_{1}(R) = -{1 \over \sqrt{2}R} BJ_{2}B^{-1} =
\pmatrix{0&0&-i/R^{2}&i \cr 0&0&0&0 \cr i&0&0&0\cr i/R^{2}&0&0&0} ,
$$
\begin{equation}\label{w1w2}
W_{2}(R) = {1 \over \sqrt{2} R} BJ_{1}B^{-1} =
\pmatrix{0&0&0&0 \cr 0&0&- i/R^{2}&i \cr 0&i&0&0 \cr 0&i/R^{2}&0&0}  ,
\end{equation}
then $W_{1}(R)$ and $W_{2}(R)$ become $F_{1}$ and $F_{2}$ of
Eq.(\ref{f1f2}) respectively in the large-$R$ limit.

The algebra given in this section is identical with that of
Sec.~\ref{3dim} based on the three-dimensional geometry of a sphere
going through a contraction/expansion of the $z$ axis.  Therefore, it
is possible to give a concrete geometrical picture to the little groups
of the Poincar\'e group governing the internal space-time symmetries
of relativistic particles.

The most general form of the transformation matrix is
\begin{equation}
D(\xi, \eta, \alpha ) = D(\xi, \eta, 0)D(0, 0, \alpha) ,
\end{equation}
with
\begin{equation}
D(\xi,\eta,0) = \exp{\left(-i\xi F_{1} - i\eta F_{2} \right)} ,
\qquad D(0,0,\alpha) = \exp{\left(-i\alpha J_{3}\right)} .
\end{equation}
The matrix $D(0, 0, \alpha)$ represents rotations around the $z$ axis
and takes the form
\begin{equation}\label{heli}
D(0, 0,\alpha) = \pmatrix{\cos\alpha & -\sin\alpha & 0 & 0 \cr
\sin\alpha & \cos\alpha & 0 & 0 \cr 0&0&1&0
\cr 0&0&0&1} .
\end{equation}
In the light-cone coordinate system, $D(\xi,\eta, 0)$ takes the form of
Eq.(\ref{compli}):
\begin{equation}\label{d12}
D(\xi, \eta, 0) = \pmatrix{1&0&0&\xi \cr0 & 1 & 0 & \eta \cr
\xi & \eta & 1 & (\xi ^{2} + \eta ^{2})/2 \cr0&0&0&1}  .
\end{equation}
This form is identical to the four-by-four matrix given in
Eq.(\ref{compli}), and therefore performs both the Euclidean and
cylindrical transformations.

\section{Four-Vectors and Gauge Transformations}\label{cyl}
Let us consider a particle represented by a four-vector:
\begin{equation}
A^{\mu }(x) = A^{\mu } e^{i(kz - \omega t)} ,
\end{equation}
where $A^{\mu } = (A_{1}, A_{2}, A_{3}, A_{0})$.  This is not a massless
particle.  In the light-cone coordinate system,
\begin{equation}
A^{\mu} = \left(A_{1}, A_{2}, A_{u}, A_{v}\right) ,
\end{equation}
where $A_{u} = (A_{3} + A_{0})/\sqrt{2}$, and $A_{v} = (A_{0} -
A_{3})/\sqrt{2}$.  If it is boosted by the matrix of Eq.(\ref{boost}),
then
\begin{equation}\label{4vec}
A'^{\mu} = \left(A_{1}, A_{2}, RA_{u}, A_{v}/R\right) .
\end{equation}
Thus the fourth component will vanish in the large-$R$ limit, while the
third component becomes large.

The momentum-energy four-vector in the light-cone coordinate system is
\begin{equation}
P^{\mu} = \left(0, 0, (k + \omega )/\sqrt{2},
(\omega - k)/\sqrt{2}\right) ,
\end{equation}
which in the rest frame becomes
\begin{equation}
P^{\mu} = \left(0, 0, m/\sqrt{2}, m/\sqrt{2}\right) ,
\end{equation}
where $m$ is the mass.  If we boost this four-momentum using
the matrix of Eq.(\ref{boost}), then
\begin{equation}
P'^{\mu} = \left(0, 0, Rm/\sqrt{2}, m/\sqrt{2}R \right) .
\end{equation}
Here again, the fourth component becomes vanishingly small for
large values of $R$, while the third component becomes large.  We
can transform the above four-momentum to that of a massless
particle by ignoring the fourth component and renormalizing the
third component, as we did for the matrices given in Eq.(\ref{w1w2}),
the above four vector can be written as
\begin{equation}
P'^{\mu} = \left(0, 0, \sqrt{2}\omega, 0 \right) ,
\end{equation}
with $R = \sqrt{2}\omega/m$.  We first took the limit of large $R$ to
eliminate the fourth component.  We then brought $R$ to a finite value
to make the third component finite.

\begin{table}
\begin{center}
\caption{Applications of group contraction to special relativity.
Massive and massless particles have different energy-momentum relations.
Einstein's special relativity gives one relation for both.  Wigner's
little group unifies the internal space-time symmetries for massive and
massless particles which are locally isomorphic to $O(3)$ and $E(2)$
respectively.}\label{jokbo}

\vspace{1mm}
\begin{tabular}{cccc}

{}&{}&{}&{}\\   
{}&{}&{}&{}\\
{} & Massive, Slow \hspace*{1mm} & COVARIANCE \hspace*{1mm}&
Massless, Fast \\[2mm]\hline
{}&{}&{}&{}\\
Energy- & {}  & Einstein's & {} \\[-0.2mm]
Momentum & $E = p^{2}/2m$ & $ E = \sqrt{p^{2}+ m^{2}} $ &
$E = p$ \\[4mm]\hline
{}&{}&{}\\
Internal & $S_{3}$ & {}  &  $S_{3}$ \\[-0.2mm]
space-time &{} & Wigner's  & {} \\[-0.2mm]
symmetry & $S_{1}, S_{2}$ & Little Group &
Gauge Transformations\\
{}&{}&{}\\[2mm]   
\end{tabular}
\end{center}
\end{table}

We can follow the same procedure to obtain the four-potential without
the fourth component:
\begin{equation}
A^{\mu} = \left(A_{1}, A_{2}, A_{u}, 0 \right) ,
\end{equation}
This process is the same as imposing the Lorentz condition
\begin{equation}
\partial^{\mu} A_{\mu}(x) = 0
\end{equation}
on the four-potential.

Let us apply the transformations of Eq.(\ref{d12}) and Eq.(\ref{heli})
to the above four-vectors.  By definition of the little group, they
do not change the four-momentum.  As for the four-potential, the
rotation matrix does not change the third and fourth component.
If we apply $D(\xi, \eta, 0)$ to the four-potential satisfying the
Lorentz condition,
\begin{equation}\label{4.9}
\pmatrix{1&0&0&\xi \cr0&1&0&\eta \cr \xi &\eta &1&(\xi ^{2} + \eta^{2})/2
\cr0&0&0&1} \pmatrix{A_{1} \cr A_{2} \cr A_{u} \cr 0}
= \pmatrix{1&0&0&0 \cr 0&1&0&0 \cr \xi & \eta &1&0 \cr 0&0&0&1}
\pmatrix{A_{1}\cr A_{2} \cr A_{u} \cr 0} .
\end{equation}
This means that transformations of the little group can be simplified
to a three-by-three matrix formalism, with the three-component column
vector $\left(A_{1}, A_{2}, A_{u} \right)$. If $D(0, 0, \alpha)$ is
applied to this vector,
\begin{equation}
\pmatrix{\cos\alpha & -\sin\alpha & 0 \cr
\sin\alpha & \cos\alpha & 0 \cr 0 & 0 & 1}
\pmatrix{A_{1} \cr A_{2} \cr A_{u}} =
\pmatrix{A_{1}\cos\alpha - A_{2}\sin\alpha
\cr A_{1}\sin\alpha + A_{2}\cos\alpha \cr A_{u}} ,
\end{equation}
which performs a rotation in the transverse plane.  If
$D(\xi, \eta, 0)$ is applied to the three-component vector,
\begin{equation}\label{3gauge}
\pmatrix{1&0&0 \cr 0 & 1 & 0 \cr \xi & \eta & 1 }
\pmatrix{A_{1} \cr A_{2} \cr A_{u}} =
\pmatrix{A_{1} \cr A_{2} \cr A_{u} + \xi A_{1} + \eta A_{2}} .
\end{equation}
The above three-by-three matrix takes the same form as the
cylindrical transformation matrix given in Eq.(\ref{cyltrans}).
This transformation does not change the transverse components, but
changes the value of the third component.  Indeed, the matrices
$D(0, 0, \alpha)$ and $D(\xi, \eta, 0)$ perform cylindrical
transformations.

Let us go back to Eq.(\ref{3gauge}).  This transformation does not
change the transverse component of the four-potential.  It changes
only the third component which is parallel to the momentum.  For
this reason, it performs a gauge transformation.  Therefor, we come
to the conclusion that transverse rotations become contracted to
gauge transformations.  The result is summarized in Table \ref{jokbo}.

\section{Lorentz-boosted Rotation Matrices}\label{lboost}
As we noted in Sec.~\ref{contrac}, In\"on\"u and Wigner considered
the contraction of the Lorentz group into the Galilean group.  There
they started from the transformation matrix, instead of generators,
and took the large-c limit.  In this way, we can trace the
transformation parameters during the limiting process.  Let us
follow this line of reasoning and take the high-speed limit of
the rotation matrices.

For a particle at rest, we can perform the rotation around the
$x$ axis using the matrix
\begin{equation}
\pmatrix{1 & 0 & 0 & 0 \cr 0 & \cos\theta & -\sin\theta & 0 \cr
0 & \sin\theta & \cos\theta & 0 \cr 0 & 0 & 0 & 1 } ,
\end{equation}
applicable to the coordinate $(x, y, z, t)$.  If we boost this
rotation matrix,
$$
\pmatrix{1 & 0 & 0 & 0 \cr 0 & 1 & 0 & 0 \cr
0 & 0 & \cosh\lambda & \sinh\lambda \cr
0 & 0 & \sinh\lambda & \cosh\lambda }
\pmatrix{1 & 0 & 0 & 0 \cr 0 & \cos\theta & -\sin\theta & 0 \cr
0 & \sin\theta & \cos\theta & 0 \cr 0 & 0 & 0 & 1 }
\pmatrix{1 & 0 & 0 & 0 \cr 0 & 1 & 0 & 0 \cr
0 & 0 & \cosh\lambda & - \sinh\lambda \cr
0 & 0 & - \sinh\lambda & \cosh\lambda } ,
$$
the result is
\begin{equation}
\pmatrix{1 & 0 & 0 & 0 \cr
0 & \cos\theta & -\sin\theta \cosh\lambda & \sin\theta \sinh\lambda
\cr 0 & \sin\theta\cosh\lambda &
\cos\theta - (1 - \cos\theta)\sinh^{2}\lambda &
(1 - \cos\theta)\cosh\lambda\sinh\lambda
\cr 0 & \sin\theta\sinh\lambda &
-(1 - \cos\theta)\cosh\lambda\sinh\lambda &
\cos\theta + (1 - \cos\theta)\cosh^{2}\lambda} .
\end{equation}
If we set $\sin\theta\sinh\lambda = \sqrt{2}\eta$ with a finite value
of $\eta$, the angle $\theta$ becomes very small in the large-$\lambda$
limit:
\begin{equation}\label{smalla}
\theta = 2\sqrt{2}\eta \exp{(-\lambda)} ,
\end{equation}
and the above matrix becomes
\begin{equation}\label{eta}
\pmatrix{1 & 0 & 0 & 0 \cr 0 & 1 & -\sqrt{2}\eta & \sqrt{2}\eta \cr
0 & \sqrt{2}\eta & 1 - \eta^{2} & \eta^{2} \cr
0 & \sqrt{2}\eta & -\eta^{2} & 1 + \eta^{2}} .
\end{equation}
Likewise, we start from the rotation matrix around the $y$ axis:
\begin{equation}
\pmatrix{\cos\phi & 0 & -\sin\phi & 0 \cr 0 & 1 & 0 & 0 \cr
\sin\phi & 0 & \cos\phi & 0 \cr 0 & 0 & 0 & 1 } ,
\end{equation}
and set $\sin\phi\sinh\lambda = - \sqrt{2}\xi$, the result is
\begin{equation}\label{phi}
\pmatrix{1 & 0 & -\sqrt{2}\xi & \sqrt{2}\xi \cr
0 & 1 & 0 & 0 \cr
\sqrt{2}\xi & 0 & 1 - \xi^{2} & \xi^{2} \cr
\sqrt{2}\xi & 0 & -\xi^{2} & 1 + \xi^{2}} .
\end{equation}
Both of the matrices in Eq.(\ref{eta}) and Eq.(\ref{phi}) appear
complicated, but they commute with each other.  The multiplication
of these two leads to
\begin{equation}
\pmatrix{1 & 0 & -\sqrt{2}\xi & \sqrt{2}\xi \cr
0 & 1 & -\sqrt{2}\eta & \sqrt{2}\eta \cr
\sqrt{2}\xi & \sqrt{2}\eta & 1 - (\xi^{2} + \eta^{2}) &
(\xi^{2} + \eta^{2}) \cr
\sqrt{2}\xi & \sqrt{2}\eta & -(\xi^{2} + \eta^{2}) &
1 + (\xi^{2} + \eta^{2})} .
\end{equation}
This is the most general form of the contracted transverse rotation
matrix.  This form as a component of the $E(2)$-like little group is
given in Wigner's original paper~\cite{wig39}, and in many later
papers~\cite{wein64}.  However, its complicated expression scared
away many physicists in the past.  In this report, we studied the
physics and mathematics of this impossible form.  In particular, we
studied the cylindrical symmetry contained in the above expression.

If we transform the above expression into the light-cone coordinate
system using the matrix given in Eq.(\ref{simil}), it becomes the
four-by-four $D(\xi, \eta, 0, 0)$ matrix given in Eq.(\ref{d12}).
What is new in this section is the expression for the angle given in
Eq.(\ref{smalla}).  This gives a physical interpretation for the
large-R renormalization procedure used in Eq.(\ref{w1w2}).  The
physical content is that for a finite value of gauge parameter, the
rotation angle becomes vanishingly small in the infinite-speed limit.
Since the rotation angle is vanishingly small, gauge transformations
are not observable.

\section*{Acknowledgments}
The author is grateful to Prof. Metin Arik and the Turkish Organizing
Committee for inviting him to this Workshop and for extending
hospitality while in Istanbul.  He would like to thank Prof. Erdal
In\"on\"u for telling us about the early days of group contractions
and his association with Eugene Wigner.  The author enjoyed very much
his collaboration with Prof. Sibel Baskal on contraction of the
Maxwell-type tensor.

\end{document}